# Imaging Tunable Luttinger Liquid Systems in van der Waals Heterostructures

*Hongyuan Li*[1, 2, 3, 10]*, *Ziyu Xiang*[1, 2, 3, 10], *Tianle Wang*[1, 3, 10], *Mit H. Naik*[1, 3, 10], *Woochang Kim*[1, 3], *Jiahui Nie*[1], *Shiyu Li*[1], *Zhehao Ge*[1], *Zehao He*[1], *Yunbo Ou*[4], *Rounak Banerjee*[4], *Takashi Taniguchi*[5], *Kenji Watanabe*[6], *Sefaattin Tongay*[4], *Alex Zettl*[1, 3, 7], *Steven G. Louie*[1, 3], *Michael P. Zaletel*[1]*, *Michael F. Crommie*[1, 3, 7]* and *Feng Wang*[1, 3, 7]*

[1]Department of Physics, University of California at Berkeley, Berkeley, CA, USA.

[2]Graduate Group in Applied Science and Technology, University of California at Berkeley, Berkeley, CA, USA.

[3]Materials Sciences Division, Lawrence Berkeley National Laboratory, Berkeley, CA, USA.

[4]School for Engineering of Matter, Transport and Energy, Arizona State University, Tempe, AZ, USA.

[5]International Center for Materials Nanoarchitectonics, National Institute for Materials Science, Tsukuba, Japan

[6]Research Center for Functional Materials, National Institute for Materials Science, Tsukuba, Japan

[7]Kavli Energy Nano Sciences Institute at the University of California Berkeley and the Lawrence Berkeley National Laboratory, Berkeley, CA, USA.

[8]Department of Materials Science and Engineering, Stanford University, Palo Alto, CA, USA.

[9]Stanford Institute for Materials and Energy Sciences, SLAC National Accelerator Laboratory, Menlo Park, CA, USA.

[10]These authors contributed equally: Hongyuan Li, Ziyu Xiang, Tianle Wang, and Mit H. Naik

**Abstract:** One-dimensional (1D) interacting electrons are often described as a Luttinger liquid[1-4] having properties that are intrinsically different from Fermi liquids in higher dimensions[5,6]. 1D electrons in materials systems exhibit exotic quantum phenomena that can be tuned by both intra- and inter-1D-chain electronic interactions, but their experimental characterization can be challenging. Here we demonstrate that layer-stacking domain walls (DWs) in van der Waals heterostructures form a broadly tunable Luttinger liquid system including both isolated and coupled arrays. We have imaged the evolution of DW Luttinger liquids under different interaction regimes tuned by electron density using a novel scanning tunneling microscopy (STM) technique. Single DWs at low carrier density are highly susceptible to Wigner crystallization consistent with a spin-incoherent Luttinger liquid, while at intermediate densities dimerized Wigner crystals form due to an enhanced magneto-elastic coupling. Periodic arrays of DWs exhibit an interplay between intra- and inter-chain interactions that gives rise to new quantum phases. At low electron densities inter-chain interactions are dominant and induce a 2D electron crystal composed of phased-locked 1D Wigner crystal in a staggered configuration. Increased electron density causes intra-chain fluctuation potentials to dominate, leading to an electronic smectic liquid crystal phase where electrons are ordered with algebraical correlation decay along the chain direction but disordered between chains. Our work shows that layer-stacking DWs in 2D heterostructures offers new opportunities to explore Luttinger liquid physics.

Landau's Fermi liquid theory has been highly successful in describing interacting electrons in two and three dimensions using concepts based on Fermionic quasiparticle excitations[5,6]. However, this picture fails for interacting electrons confined to one dimension (1D) and leads to Luttinger liquid behavior[1-4], where the elementary excitations are bosonic. The 1D electron interaction strength and resulting Luttinger liquid behavior can be tuned continuously by varying the electron density. Isolated 1D electron chains at high electron density are well described by a weakly-interacting Luttinger liquid theory[2-4] (i.e. where electron interactions are relatively weak). This description features spin-charge separation where low-energy excitations are described by plasmons and spinons. In the low-density limit, however, 1D electrons cross over into a regime that features a high susceptibility to quasi-long-range Wigner crystallization due to stronger electron interactions. This regime has a qualitatively different picture for spin-charge separation, where the charge mode is described as an electron crystal phonon and the spin mode is determined by antiferromagnetic exchange coupling between nearest-neighbor electrons[7,8]. Here spin exchange interactions are highly suppressed and thermal fluctuations dominate, making the system a *spin-incoherent* Luttinger liquid[7,8]. For intermediate electron densities the spin exchange interaction can lead to an unusual form of magneto-elastic coupling[7,8] where the electron lattice dimerizes and forms valence-bond spin singlet pairs to lower the overall magnetic energy (i.e., the spin-Peierls effect), analogous to the Su-Schrieffer-Heeger model. The electron crystal ultimately evolves into a standard linear Luttinger liquid as spin and charge energies become comparable at sufficiently high electron density. Because crystalline order is algebraic in 1D, the evolution from low to high density is a crossover. Arrays of 1D electron chains host even richer phenomena due to the interplay between intra- and inter-chain interactions. Depending on the strength of inter-chain interactions many new quantum phases have been predicted theoretically, including 2D electron crystals[9,10], electron smectic liquid crystals[10], and even sliding Luttinger liquids exhibiting non-Abelian fractional quantum Hall states[11,12].

In the last few decades strong efforts has been made to experimentally explore Luttinger liquids. Weakly interacting Luttinger liquids have been observed in 1D metals[13,14], semiconductor nanowires[15-18], topological edges states[19-22], and twin-boundary defects[23,24], where spin-charge separation and power-law scaling of tunneling probability have been observed. Characterizing strongly interacting 1D electrons at lower densities is more difficult since they are sensitive to inevitable weak disorder and stray fields. Suspended semiconducting carbon nanotubes have provided a useful platform to explore the low-density regime and signatures of Wigner crystallization have been observed in carbon nanotube electrical transport[25] and scanning single-electron transistor (SET) measurements[15]. However, even few-electron Wigner crystals in these nanotubes are strongly distorted by disorder, thus preventing study of quasi-long-range order and the crossover from a strongly interacting Wigner crystal to a weakly interacting Luttinger liquid. Experimentally characterizing arrays of coupled Luttinger liquids is even more challenging for lack of a suitable platform. It has been suggested that the stripe phase of high temperature superconductors[26] and the anisotropic moire superlattice in twisted $WTe_2$[27] might provide coupled 1D electron chains, but microscopic descriptions of these materials are still lacking.

Here we demonstrate that layer-stacking domain walls (DWs) in bilayer $WS_2$ form an ideal platform for exploring spin and orbital quantum behavior in 1D Luttinger liquids with tunable interaction strength. Stacking DWs can be formed either in isolated form (yielding single 1D electron chains) or as self-assembled periodic arrays of Luttinger liquids. An advantage of DWs is that they are embedded in two-dimensional (2D) van der Waals heterostructures that exhibit low structural disorder and facilitate convenient electrical device fabrication and characterization. Using a novel scanning tunneling microscopy (STM) technique we have directly imaged the evolution of DW-based Luttinger liquids under different interaction regimes that reveal new quantum phenomena. We find that isolated DWs exhibit almost perfect 1D Wigner crystals pinned by dilute defects at low electron density. In this regime density matrix renormalization group (DMRG) calculations suggest that exponentially suppressed spin interactions are dominated by thermal excitations for our experimental temperature, giving rise to spin-incoherent Luttinger liquid behavior. At increased electron densities we experimentally observe dimerized Wigner crystals that are consistent with theoretical predictions of enhanced magneto-elastic coupling between an antiferromagnetic spin chain and the electron charge lattice. Our DMRG calculations in this regime suggest that this dimerization is associated with enhanced susceptibility to oscillating valence bond order. At even higher density the observed Wigner crystals evolve into weak-interacting linear Luttinger liquids both experimentally and in our calculations. For periodic arrays of DWs in the low electron density regime, we experimentally observe anisotropic 2D electron lattice behavior arising from phase-locked 1D Wigner crystals. At increased densities a new electron smectic liquid crystal phase emerges.

Our experimental setup involves an artificially stacked 60°-twisted bilayer $WS_2$ device integrated into a scanning tunneling microscope (STM) as shown in Fig. 1a (see methods for fabrication details). The artificial stacking technique introduces small angle variations and strain that generate stacking DWs in the bilayer $WS_2$. The bilayer $WS_2$ is placed on top of a hBN flake with thickness $d_{hBN} = 67$nm that is placed above a graphite back gate. A back gate voltage $V_{BG}$ is applied to electrostatically dope electrons into the bilayer $WS_2$. The sample-tip bias $V_{bias}$ applied between the WS$_2$ and the STM tip is chosen to achieve two goals simultaneously: (1) alignment of the tip chemical potential $\mu_{tip}$ within the WS$_2$ bandgap to allow electrons tunneling from the (electron-doped) WS$_2$ conduction band edge to the tip and (2) alignment of the tip and WS$_2$ vacuum energy levels to ensure minimal tip perturbation of the surface. These conditions enable us to directly map the 1D interacting electron density. We employ a graphene nanoribbon (GNR) contact electrode on the WS$_2$ surface to minimize the device resistance and facilitate STM measurement[28].

Our STM topographic images of bilayer WS$_2$ (Fig. 1b) reveal a series of 1D structures corresponding to stacking domain walls (DWs). A single DW separates two AB stacking regions with an in-plane dislocations of one unit cell characterized by a Burgers vector[29-35]. We observe different DWs configurations, including isolated DWs (Fig. 1b top left) and few-DW clusters (Fig. 1b bottom left). DWs are also observed to self-assemble into periodic arrays (Fig. 1b right) with inter-DW separation of $L_{DW} \approx 8.2$nm. The DW atomic structure depends on the angle $\theta$ between the DW direction and the Burgers vector. Two extreme configurations are for shear ($\theta = 0$) and tensile ($\theta = \frac{\pi}{2}$) DWs[29,30]. The shear DW atomic structure is illustrated in Fig. 1c and shows a DW separating two AB stacking regions. Here a guide to the eye shows the W atoms in the bottom layer (orange dots) and the S atoms in the top layer (blue dots) across the DW to help visualize the interlayer dislocation. While the S and W atoms are on top of one another to the left of the DW, they are separated by a unit cell to the right of the DW. Most DWs observed in this work have a Burgers vector angle $\theta$ between 0 and $\frac{\pi}{2}$, as identified through surface topography and atomically resolved STM images (for example see Fig. S1).

Combining density functional theory (DFT) calculations with our STM spectroscopy reveals that the conduction band minimum in a DW is lower than for an AB stacked region (see SI section 3 and 4 for details). This causes electrostatically doped electrons (controlled via the back gate voltage) to be confined within the DWs and thus provides a platform for studying Luttinger liquids. We are able to directly image the electron distribution in DWs by measuring the tunnel current from DW conduction band edge (CBE) states[36], as sketched in Fig. 2a. Here the chemical potential of electron-doped $WS_2$ ($\mu_{tip}$) lies above the CBE so that when the tip chemical potential is lowered into the $WS_2$ band gap then the tunnel current (denoted CBE tunnel current) comes from the doped $WS_2$ electrons. The tip bias voltage is tuned to minimize tip-induced perturbation to electrons in the DWs by finding $V_{bias}$ that aligns the sample and tip vacuum levels and yields the best imaging quality (see SI section 5 for details). The CBE tunnel current map directly reflects the spatial distribution of doped electrons as demonstrated previously in CBE current measurements on Wigner molecular crystals in moire artificial atoms[36]. Fig. 2b shows the tunnel current I-V characteristic on a log scale as a function of $V_{BG}$ measured at the center of a DW with a large tip-sample separation (determined by the setpoint condition of $V_{bias}$ = -3.30V, $I_{sp}$ = 20pA, and $h_{tip}$ = -50pm, see methods for details). Negligible tunnel current occurs for -1.8V < $V_{bias}$ < 0, which corresponds to the $WS_2$ semiconducting band gap (the valence and conduction band edges are marked with white dashed lines). The CBE tunnel current is lower than the measurement noise floor in the bandgap region due to the large tip-sample separation. When the tip height is slightly lowered (setpoint condition: $V_{bias}$ = -2.70V, $I_{sp}$ = 20pA, and $h_{tip}$ = -100pm) then the CBE tunnel current start showing up in the gap region, as seen in Fig. 2c. Here dispersing features appear in the range 3V < $V_{BG}$ < 8V that correspond to quantized electron number changes in a finite-length 1D Wigner crystal (see additional details in SI section 10).

The positions of the electrons in a DW can spatially mapped by through the CBE tunnel current. Fig. 2d shows the evolution of the CBE tunnel current map for a DW (central one in triple DWs) in the low electron density regime for $2.5V < V_{BG} < 9.0V$. Three bright spots labeled as A, B, C remain fixed even as the electron density is changed and are attributed to three separate defects in the DW that each pin an electron. Between defects B and C we observe a periodic lattice of highly localized electrons with a quantized electron number that increases from 4 to over 15 as $V_{BG}$ is increased from 2.5V to 9.0V. This tunable 1D electron lattice provides direct evidence for 1D Wigner crystal formation in DWs. Although a true crystal with long-range order is theoretically forbidden in an infinite 1D system, a well-defined Wigner crystal with quasi-long-range order can be stabilized by long range Coulomb interactions in a finite 1D chain[37], as observed here.

We note that the observed 1D electron lattices show almost no disorder except for the defects pinning the ends of the Wigner crystal chain, thus indicating the high quality of the 1D domain wall system. The Wigner crystal persists to a surprisingly high electron density. We can characterize the electron density using the dimensionless parameter $r_s = \frac{d}{a_B}$ where $d$ is the electron separation, $a_B = \frac{4\pi\varepsilon\hbar^2}{m_e e^2}$ is the effective Bohn radius, $m_e = 0.39 m_0$ is the effective electron mass obtained from our DFT calculations and $\varepsilon = 3.9\varepsilon_0$ is the relative permittivity ($\varepsilon_0$ is the vacuum permittivity). The values of $d$ and $r_s$ for the 1D lattice shown in Fig. 2d are listed in the table of Fig. 2e and stand in stark contrast to the case of 2D electron gases where the Wigner crystal state only exists for $r_s > 30$[38-40]. We see that the 1D Wigner crystal state is present for our DW lengths even at $r_s < 8$. This is a signature of the strong impact that electron-electron interactions have in 1D.

At increased electron densities, a crossover from the 1D Wigner crystal to a dimerized Wigner crystal and then to a weakly interacting Luttinger liquid is observed. Fig. 3a shows the evolution of the CBE tunnel current map for 1D electrons in a DW containing a single defect over the range $8.5V < V_{BG} < 14.5V$. For better visualization of the experimental data Fig. 3b shows a 2D plot of the normalized current as a function of $V_{BG}$ and horizontal position where each horizontal line is obtained by vertically averaging the pixels for each image in Fig. 3a. Wigner crystal formation is clearly seen at low electron densities, but as $V_{BG}$ is raised above 10V an unexpected distortion of the Wigner crystal is observed where adjacent peaks merge and form dimers. The dimerization becomes more pronounced at increased electron density until $V_{BG}$ reaches ~11.5V, at which point the two peaks in each dimer pair merge into a single broad peak. For $V_{BG} >11.5V$ the Wigner crystal is gone and a new pattern with doubled period emerges. This new pattern corresponds to the charge density oscillation (i.e., Friedel oscillation) of a Luttinger liquid in the weakly-interacting regime.

The crossover from Wigner crystal to weakly interacting Luttinger liquid can also be seen in momentum space. Fig. 3c shows the fast Fourier transformation (FFT) of Fig. 3b and exhibits two distinct wavevector peaks at different electron densities. In a non-interacting two-fold degenerate 1D system (i.e., due to spin-valley locking) the Fermi wavevector is $k_F = \frac{\pi N}{2L}$, where $L$ is the 1D chain length and $N$ is the total electron number. The Friedel oscillation arising from the backscattering of electrons then has a wavevector $k_{Fr} = 2k_F = \frac{\pi N}{L}$. The Wigner crystal state, on the other hand, has a period of $L/N$ and thus features a wavevector $k_W = \frac{2\pi N}{L} = 4k_F$. The two dispersive peaks in Fig. 3c are therefore assigned to the Wigner crystal (red dashed line) and Friedel oscillation (white dashed line). The Wigner crystal branch (4k$_F$) is seen to become weaker with increased electron density and to disappear at $V_{BG} \approx 11.5V$. The Friedel oscillation branch (2k$_F$) emerges at $V_{BG} \approx 10V$ and becomes the dominant periodicity gradually with the increased electron density. Both branches are seen in the intermediate region $10V < V_{BG} < 11.5V$ which corresponds to the dimerized Wigner crystal regime. Extended data displaying the crossover from Wigner crystal to weakly interacting Luttinger liquid are included in the SI section 5.

In order to better understand the different density-dependent regimes observed experimentally for isolated DWs, we performed numerical density matrix renormalization group (DMRG) calculations (see SI section 6 for details). Fig. 3d shows the resulting local electron density profile as a function of average density for a DMRG calculation of a finite 1D chain of electrons at T = 0. The corresponding FFT plot is shown in Fig. 3e. Comparison of Figs. 3d, e with Figs. 3b, c shows that the DMRG calculation captures the main features of the experiment including the dimerization of the Wigner crystal with increasing electron density followed by a crossover into the weakly interacting Luttinger liquid regime featuring a 2k$_F$ oscillation.

The DMRG result shows the 2k$_F$ oscillation (Fig. 3e) extending further into the low-density regime than is seen experimentally (Fig. 3c). This likely originates from the thermal excitation of the spin degree of freedom during the experimental measurement. We can gain insight into this process by considering the energetics of a 1D chain of electrons compared with the thermal background. At lower densities the electrons are far apart with reduced wavefunction overlap and so the spin exchange energy is strongly suppressed. This is reflected in the DMRG calculated spin excitation energy E$_J$ (i.e., the energy to flip a single spin in an 80nm long electron chain (details can be seen in SI section 7)) which is plotted in Fig. 3f and shows exponential reduction with separation for n < 0.30nm$^{-1}$. This is much smaller than the charge excitation energy $\hbar\omega_0$ (i.e., the zero-momentum longitudinal optical phonon energy of a 1D Wigner crystal (details can be seen in SI section 7)). For n < 0.30nm$^{-1}$ the thermal energy $k_B T$ at T = 5.4K dominates over the spin energy, thus precluding any possible spin coherence. This regime has been theoretically explored previously and is referred to as a spin-incoherent Luttinger liquid[7,8] where thermally induced spin incoherence suppresses the 2k$_F$ Friedel oscillation relative to the 4k$_F$ Wigner crystal[7,8], consistent with our experimental data.

The dimerization phenomenon occurring at slightly higher density (0.30nm$^{-1}$ < n < 0.38nm$^{-1}$) arises when the spin exchange energy E$_J$ exceeds the thermal energy $k_B T$, thus making the spin behavior important in the Wigner crystal. In this regime the Wigner crystal electrons are antiferromagnetically coupled and the exchange energy $E_J$ strongly depends on the electron separation $d$ (Fig. 3g). This causes magneto-elastic coupling between the spin and charge of the Wigner crystal, as described by the expression[7]

$$H_{s-c} = J_1 \sum_l (u_{l+1} - u_l) \mathbf{S}_{l+1} \cdot \mathbf{S}_l.$$

Here $J_1 = \partial J/\partial d$ where $J$ is the spin exchange interaction (which is related to E$_J$, see SI section 7), and $u_l$ and $\mathbf{S}_l$ are the spatial coordinate and spin of the $l$th electron. This spin-lattice interaction term is similar to what is seen in spin-Peierls systems[41,42], except that the phonon originates from an electron lattice rather than an atomic lattice. Magneto-elastic coupling lowers the overall energy of the 1D system by dimerizing the lattice to gain magnetic energy (illustrated in Fig. 3g), thus enhancing the 2k$_F$ density oscillation as $J_1$ increases at higher electron density. This order is expected to decay algebraically from the defects pinning the crystal.

Evidence for the magnetic origin of the dimerization phenomenon can be further seen by calculating the entanglement entropy $S_{EE}$ of the 1D electron chain. $S_{EE}$ reflects the degree of entanglement for adjacent electrons and so provides a measure of spin singlet formation. Fig. 3h shows the DMRG calculation of entanglement entropy $S_{EE}$ across each site of the electron chain at n = 0.3 nm$^{-1}$ (see SI section 6 for details). A clear oscillation in the entanglement entropy is observed, suggesting a high degree of entanglement between each dimerized electron pair and thus a tendency toward spin singlet formation. The combination of spin-lattice coupling and alternating valence bond order in the dimerized 1D Wigner crystal is analogous to the physics of 1D polyacetylene as described by the Su–Schrieffer–Heeger (SSH) model[43].

In the high-density regime (n > 0.38nm$^{-1}$)) the electron kinetic energy overcomes the Coulomb interaction, and the system behaves as a weakly interacting Luttinger liquid. Here the electron density is dominated by the 2$k_F$ Friedel oscillation and the 4$k_F$ oscillation gradually vanishes due to delocalization of the Wigner crystal.

The interactions *between* 1D electron chains in DW arrays lies beyond the scope our 1D DMRG treatment, but this is also a regime where we observe rich experimental phenomenology, including new quantum phases arising from the interplay between intra- and inter-DW interactions. Figs. 4a-4h show the experimental evolution of the CBE tunnel current for an array of DWs as $V_{BG}$ is increased over the range 2.0V < $V_{BG}$ <5.0V. We find that the distribution of electrons across the DWs changed dramatically as the overall density is changed. At low electron density (Figs. 4a-4d) the DW Wigner crystal chains exhibit a staggered structure with electrons in one chain aligning with empty sites in the neighboring chain. This can be seen in the blue arrows of Fig. 4c which trace the zigzag path between electrons in neighboring chains (defects are identified by yellow circles in Fig. 4g). The staggered structure minimizes inter-DW interactions (i.e., by maximizing the separation between electrons in neighboring DWs) thus producing a new anisotropic 2D electronic crystalline phase. Figs 4i-4l show 2D FFT plots of the low-density staggered phase shown in Figs. 4a-4d. Sharp diffraction peaks charactering this new crystalline phase can be clearly observed.

At higher electron density (Figs. 4e-4h) the staggered electronic phase dissociates into a new electronic configuration. This can be seen in a trace of neighboring electrons for $V_{BG} = 4.0V$ which displays an almost random path (Fig. 4f, blue arrows). In this regime a pinned 1D Wigner crystal remains in each DW but inter-DW coherence vanishes, similar to the transition from 2D crystalline state to 2D smectic state seen in liquid crystals. A smectic liquid crystal-like phase is confirmed by the 2D FFT plots of Figs. 4m-4p which exhibit "trivial" peaks due to the periodic DW line array (marked with red circles) as well as "nontrivial" feature (marked by blue ovals) that arise due to the interior 1D Wigner crystal periodicity. Intensity within each blue oval reflects a constant wavevector *along* the DW direction (corresponding to the 1D Wigner crystal periodicity), but the diffuseness of the feature reflects randomness in the direction perpendicular to the DW and thus indicates disorder between different Wigner crystal chains. Such diffraction patterns are characteristic of more conventional smectic liquid crystal phases[10,44,45] (additional data can be seen in section 8).

This 2D crystalline-to-smectic transition originates from the interplay between intra- and inter-DW interactions, as well as defect-induced intra-DW potential fluctuations. At low electron densities we expect the inter-DW interaction ($E_{inter}$) and intra-DW interaction ($E_{intra}$) to both be higher than the defect-induced fluctuation potential ($E_D$). Together they stabilize the anisotropic 2D electron lattice composed of phase-locked 1D Wigner crystals that exhibit a staggered electron configuration with weak distortions to accommodate disorder. For increased electron density, however, the inter-DW interaction is expected to rapidly decrease as $E_{inter} \sim e^{-\sqrt{2}\pi nl}$, where $n = N/L$ is the 1D chain electron density and $l$ is the distance between adjacent DWs (see SI section 9 for details). Here a new energy hierarchy emerges where $E_{intra} > E_D > E_{inter}$, causing the Wigner crystal within each chain to remain stabilized while inter-chain coherence is destroyed by disorder.

In conclusion, we show that layer-stacking DWs arising from differential uniaxial strain in van der Waals heterostructures offer tremendous opportunities to explore Luttinger liquid physics. Although we used the simple 2D semiconductor $WS_2$ as a model system here, similar isolated DWs and periodic DW arrays can be realized in any 2D bilayers with uniaxial heterostrain. A wide variety of exotic Luttinger liquid phenomena could emerge from DWs in novel van der Waals heterostructures, such as 2D charge density wave materials, 2D magnets, and 2D superconductors.

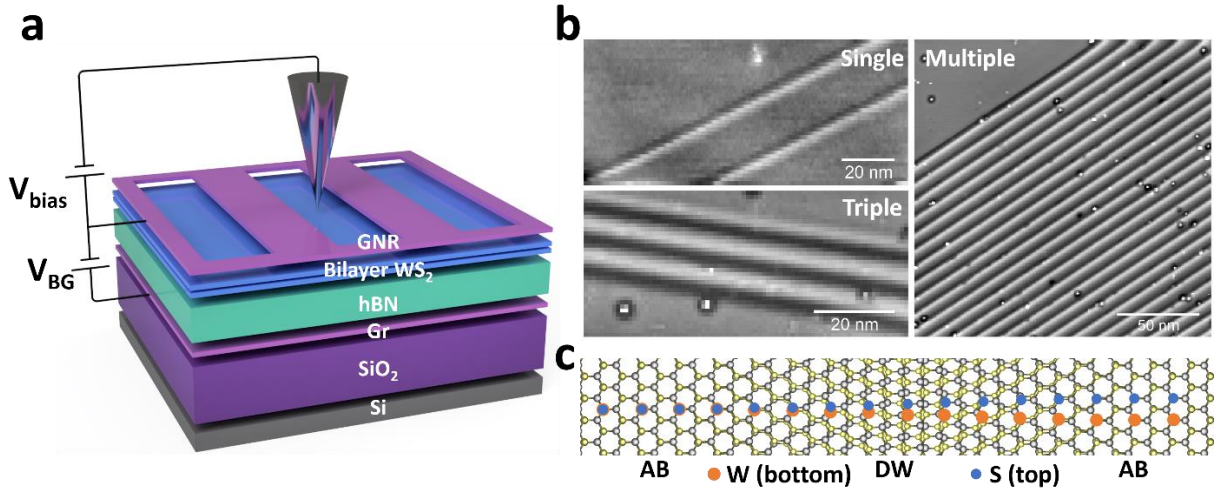

**Figure 1. Stacking domain walls in a bilayer WS$_2$. a**. Schematic of the STM measurement of a gate-tunable bilayer WS$_2$ device. The bilayer WS$_2$ is placed on top of a 67nm thick hBN layer and a graphite substrate that defines the back gate. A back gate voltage V$_{BG}$ is applied to electron-dope the WS$_2$. A sample-tip bias V$_{bias}$ is applied to the WS$_2$ and relative to the STM tip. A graphene nanoribbon (GNR) array is placed on top of the WS$_2$ to serve as the contact electrode. **b**. Typical STM topographic images of the stacking domain walls (DWs) in the bilayer WS$_2$ that separates two AB stacking regions with a one-unit-vector interlayer dislocation. The DWs exhibit different aggregation behavior including single DWs (left top), triple DWs (left bottom) and DW periodic arrays (right). The inter-DW distance in the triple DWs and DW arrays is ~ 8nm. **c**. Sketch of the atomic structure for a shear-type stacking DW. The left and right regions are AB stacked while the center shows a vertically aligned DW. The positions of W atoms in the bottom layer (orange dots) and S atoms in the top layer (blue dots) are highlighted along a linecut across the DW. For shear-type DWs, the two AB stacking regions have an interlayer unit-vector shift parallel to the DW.

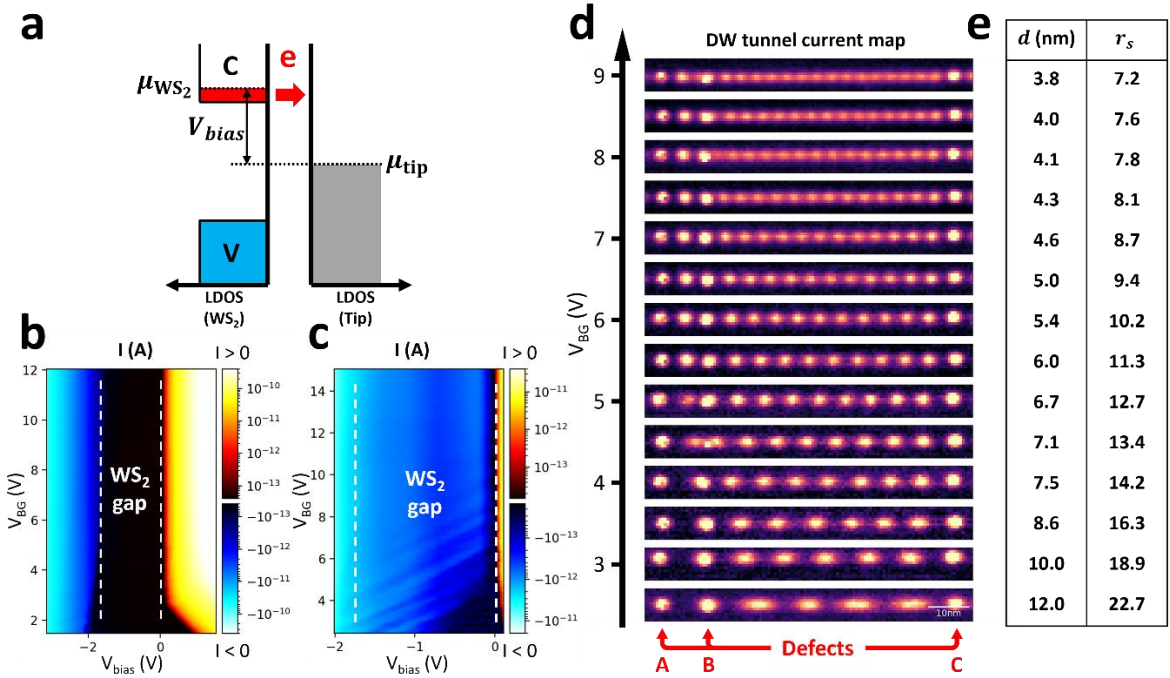

**Figure 2. Tunnel current measurement of 1D Wigner crystal. a**. Schematic energy diagram for the conduction band edge (CBE) tunnel current measurement of electron-doped WS$_2$. The WS$_2$ chemical potential $\mu_{WS_2}$ lies above the CBE. When the tip chemical potential $\mu_{tip}$ (controlled by V$_{bias}$) is the aligned within the band gap of the WS$_2$, the tunnel current arises from the doped electrons at the conduction band edge. **b,c.** Tunnel current I-V characteristics as a function of V$_{BG}$ measured at the DW center for electron-doped WS$_2$ with (**b**) a large (V$_{bias}$ = -3.30V, I$_{sp}$ = 20pA, h$_{tip}$ = -50pm) and (**c**) small (V$_{bias}$ = -2.70V, I$_{sp}$ = 20pA, h$_{tip}$ = -100pm) tip-sample separation. The current is plotted on a log scale with the positive and negative branches using different colormaps. The CBE and valence band edge (VBE) are marked with white dashed lines. For small tip-sample separation a negative CBE tunnel current can be seen in the WS$_2$ gap. **d**. CBE tunnel current maps for the center DW in a triple-DW group with V$_{BG}$ increased from 2.5V to 9.0V. V$_{bias}$ is selected in the range -0.85V < V$_{bias}$ < -0.30V to minimize the tip-sample vacuum level mismatch (tip setpoint: V$_{bias}$ = -2.70V, I$_{sp}$ = 20pA, and h$_{tip}$ = -100pm). The maps reveal a pinned 1D Wigner crystal where each bright dot corresponds to one localized electron. Three pinning defects are labeled with red arrows at the bottom. **e**. Table of the electron separations and corresponding values of r$_s$ for the images shown in (**d**).

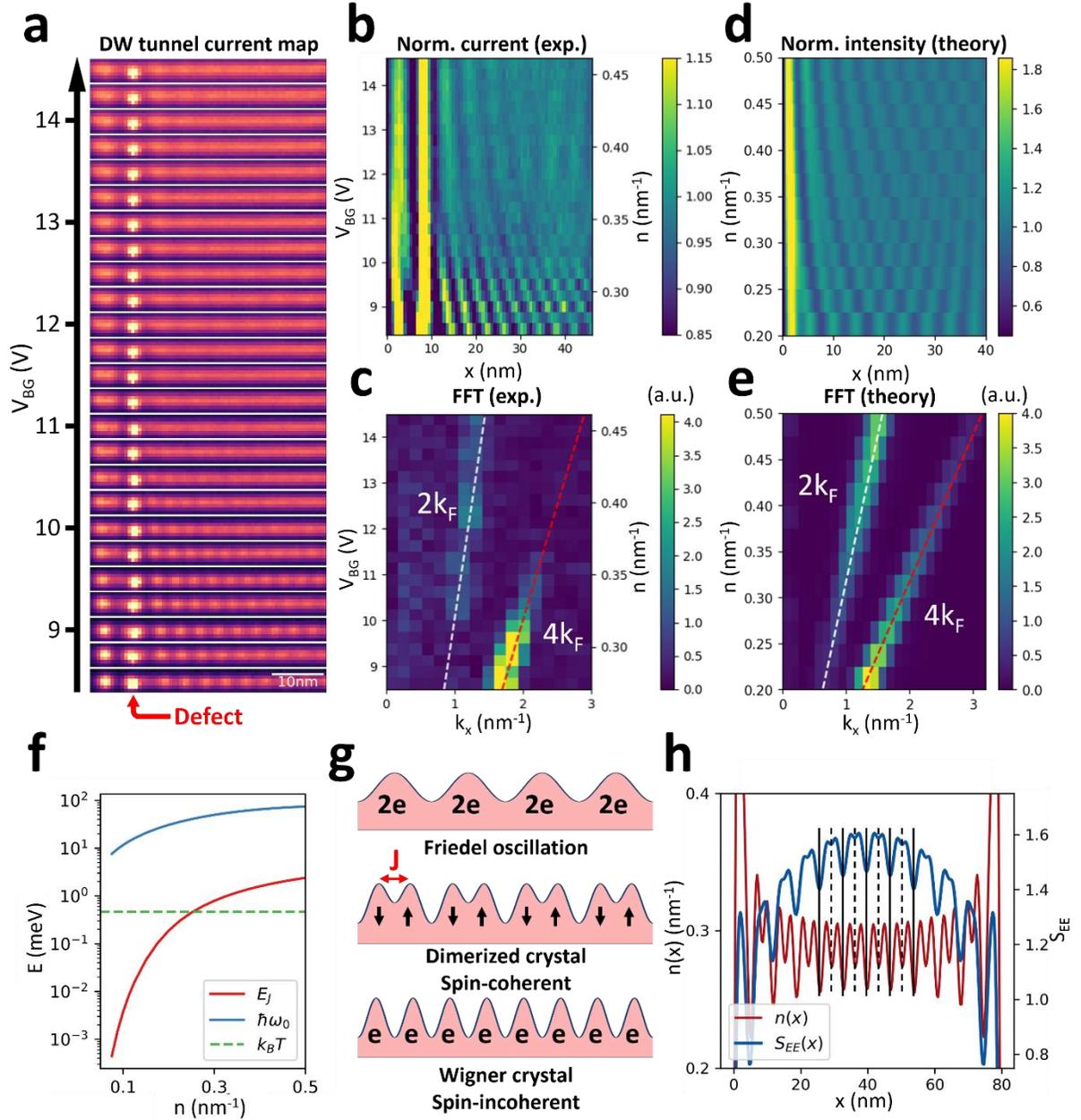

**Figure 3. 1D Wigner-Friedel crossover. a.** Evolution of CBE tunnel current maps for a center DW in a triple-DW group with $V_{BG}$ increased from 8.5V to 14.5V. The periodic structure of the 1D Wigner crystal is gradually replaced by a new periodic structure having twice the period. A defect near the left end is labeled with a red arrow. $V_{bias}$ = -0.23V (tip setpoint: $V_{bias}$ = -2.70V, $I_{sp}$ = 20pA, and $h_{tip}$ = -100pm). **b.** 2D plot of the normalized CBE tunnel current of the DW in (**a**). Each line comes from vertically averaging the pixels of an image from (**a**). **c.** Fast Fourier transformation (FFT) of the data shown in (**b**). Two dispersing peaks (labeled with dashed lines) correspond to $2k_F$ and $4k_F$ where $k_F$ is the Fermi wavevector as defined in the text. **d.** Density matrix renormalization group (DMRG) calculation of the 1D spatial charge distribution as a function of electron density n. The simulated 1D chain is 80nm long with hard-wall ends. The charge distribution is symmetric about x = 40nm and so we only show the results for x < 40nm. **e**. FFT of the result shown in (**d**). Dispersing peaks at $2k_F$ and $4k_F$ can be seen. **f**. Calculated charge phonon energy $\hbar\omega_0$ and exchange interaction energy $E_J$ as a function of 1D chain electron density. The experimental temperature energy scale is labeled with a green dashed line. **g**. Schematic illustration of 1D electron chain with decreasing interaction strength (from bottom to top) shows three regimes: Wigner crystal, dimerized crystal, and Friedel oscillation. **h**. DMRG calculation of the charge density $n(x)$ and entanglement entropy $S_{EE}$ across each site of 1D electron chain for n = 0.3 nm$^{-1}$ (see SI for details). $S_{EE}$ reflects the degree of entanglement (and therefore singlet formation) between neighboring electrons in a dimerized Wigner crystal. Vertical lines label the boundary (solid) and center (dashed) of singlet pairs.

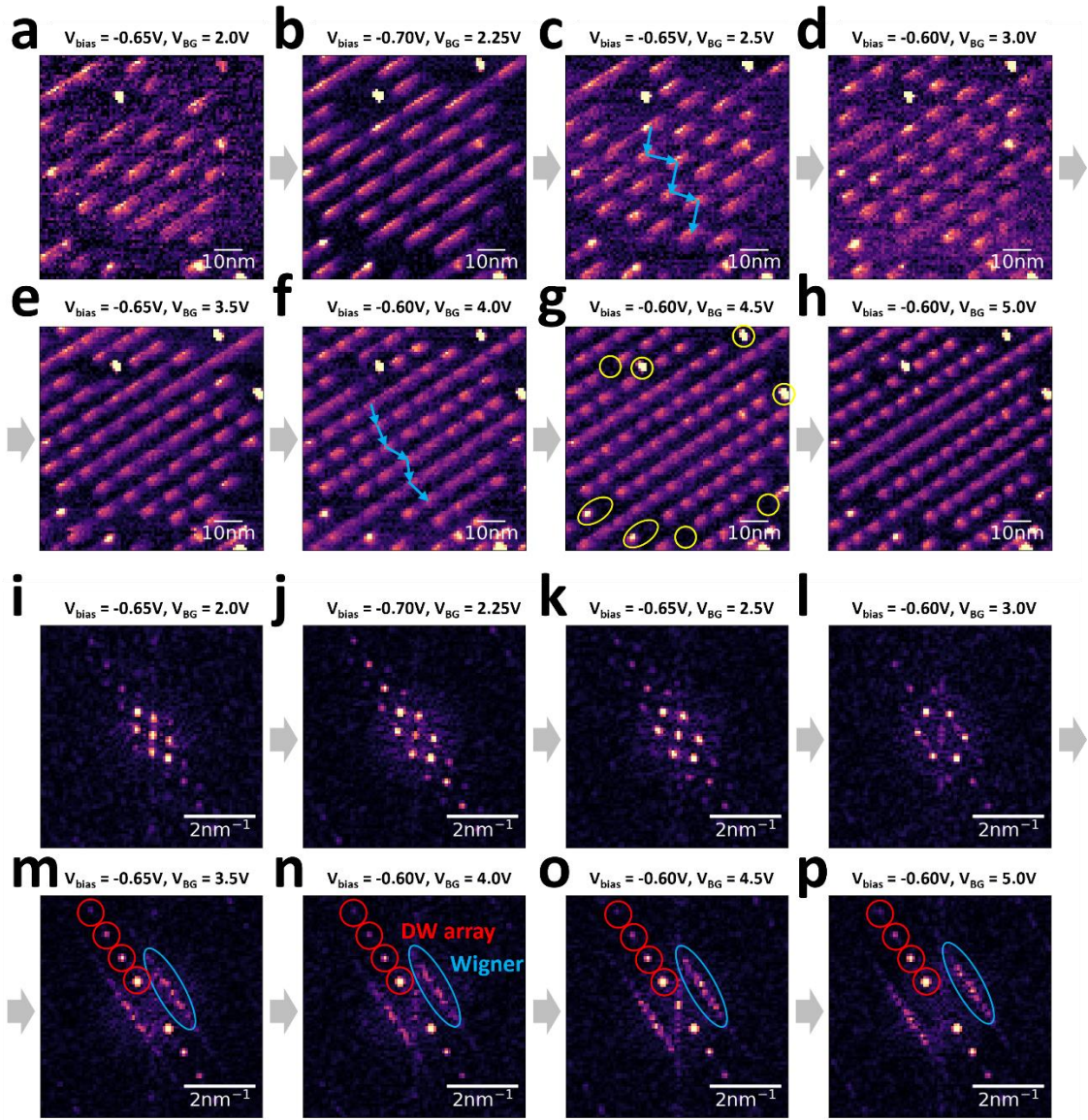

**Figure 4. Electron crystalline-to-smectic transition in 1D domain wall array. a-h.** CBE tunnel current maps for a periodic array of 1D DWs as $V_{BG}$ is increased from 2.0V to 5.0V. Defects are marked with yellow circles in (**g**). At low electron density (**a-d**) an array of 1D Wigner crystals is seen that exhibits a staggered structure, forming a new 2D crystalline phase. At high electron density (**e-h**) the staggered order is lost and spatial coherence between nearby 1D Wigner crystals disappears, forming a new electronic smectic phase. Blue arrows label the positions of electrons in nearest-neighbor DWs. **i-p**. 2D fast Fourier transform (FFT) plots of the images shown in (**a-h**). At low density (**i-l**) the staggered Wigner crystal array yields well-defined peaks in the FFTs due to the presence of 2D crystalline structure. At high density (**m-p**) two features dominate the 2D FFT plots: (i) peaks reflecting the periodic spacing of DW lines (labeled with red circles), and (ii) diffuse lines (labeled with blue ovals) that reflect 1D Wigner crystal periodicity but lack inter-DW spatial coherence (i.e., the smectic liquid crystal state).

**Corresponding Author**

* Email: hongyuan_li@berkeley.edu (H.L.), mikezaletel@berkeley.edu (M.Z.), crommie@physics.berkeley.edu (M.C.) and fengwang76@berkeley.edu (F.W.).

**Author Contributions**

H.L., M.P.Z., M.F.C., and F.W. conceived the project. H.L., Z.X., J.N., and S.L. fabricated the WS$_2$ heterostructure device. H.L. and Z.X. performed the STM/STS measurement of the WS$_2$ device. T.W. and M.P.Z. performed the DMRG calculations of the 1D interacting electrons. M.H.N., W.K., and S.G.L. performed the *ab initio* calculations of the DW structures. Z.G. and Z.H. performed the measurment of the QPI in bilayer MoSe$_2$. H.L., Z.X., A.Z., M.F.C. and F.W. discussed the experimental design and analyzed the experimental data. Y.O., R.B. and S.T. grew the WS$_2$ crystals. K.W. and T.T. grew the hBN single crystal. All authors discussed the results and wrote the manuscript.

**Notes**

The authors declare no financial competing interests.

**ACKNOWLEDGMENT**

This work was primarily funded by the U.S. Department of Energy, Office of Science, Basic Energy Sciences, Materials Sciences and Engineering Division under Contract No. DE-AC02-05-CH11231 within the van der Waals heterostructure program KCFW16 (device fabrication, STM spectroscopy, theoretical analyses and computations). Support was also provided by the National Science Foundation Award DMR-2221750 (surface preparation). This research used the Lawrencium computational cluster provided by the Lawrence Berkeley National Laboratory (Supported by the U.S. Department of Energy, Office of Basic Energy Sciences under Contract No. DE-AC02-05-CH11231). S. T. acknowledges primary support from U.S. Department of Energy-SC0020653 (materials synthesis), NSF CMMI1825594 (NMR and TEM studies on crystals), NSF DMR-1955889 (magnetic measurements on crystals), NSF ECCS2052527 (for bulk electrical tests), DMR 2111812, and CMMI 2129412 (for optical tests on bulk crystals). K.W. and T.T. acknowledge support from the JSPS KAKENHI (Grant Numbers 21H05233 and 23H02052) and World Premier International Research Center Initiative (WPI), MEXT, Japan. The work at Massachusetts Institute of Technology was supported by the Air Force Office of Scientific Research (AFOSR) under award FA9550-22-1-0432. H. L. acknowledges the support from Kavli Energy Nano Sciences Institute graduate student fellowship. The authors acknowledge the Texas Advanced Computing Center (TACC) at The University of Texas at Austin for providing high performance computing resources. This research also used resources of the National Energy Research Scientific Computing Center (NERSC), a U.S. Department of Energy Office of Science User Facility located at Lawrence Berkeley National Laboratory, operated under Contract No. DE-AC02-05CH11231.

**Methods**

The bilayer $WS_2$ device was fabricated using a micromechanical stacking technique[46]. A poly(propylene) carbonate (PPC) film stamp was used to pick up all exfoliated 2D material flakes. The 2D material layers in the main heterostructure region were picked up in the following order: substrate hBN, graphite, bottom hBN, monolayer $WS_2$, second monolayer $WS_2$ (with 60° twist), graphene nanoribbon array. The graphene nanoribbon array serves as a contact electrode for the twisted $WS_2$. The PPC film and stacked sample was peeled together, flipped over, and transferred onto a Si/SiO$_2$ substrate (SiO$_2$ thickness 285nm). The PPC layer was subsequently removed using ultrahigh vacuum annealing at 330 °C, resulting in an atomically-clean heterostructure suitable for STM measurements. 50nm Au and 5nm Cr metal layers were evaporated through a shadow mask to form electrical contacts to graphene layers. Some residues on the shadow mask are occasionally transferred to the sample surfaces during the evaporation. These residues are later cleaned by the STM tip scratching. The STM tip is further cleaned on the GNR surface through field emission and/or tip bias pulse.

**STM tunnel current and dI/dV spectroscopy measurement**: Tunnel current and dI/dV spectrum measurements were performed under open-loop conditions with the tip height first stabilized at tip bias $V_{bias}$ and setpoint current $I_{sp}$ with closed feedback and then lowered by turning off the feedback and reducing the tip height by a distance of $h_{tip}$. A bias modulation with 25mV amplitude and 500~900 Hz frequency was applied to obtain the dI/dV signal. All STM measurements were performed at T=5.4K.

**Supplementary Materials**

**Data availability**

The data supporting the findings of this study are included in the main text and in the Supplementary Information files, and are also available from the corresponding authors upon request.